\documentclass[prd,aps,twocolumn,floatfix,showpacs,nofootinbib]{revtex4-1}

\usepackage{graphicx,psfrag}
\usepackage{amsmath,amsfonts,amssymb}
\usepackage{mathrsfs}
\usepackage{multirow,hyperref}
\usepackage{comment}
\usepackage{pifont}
\usepackage[normalem]{ulem}

\newcommand{\be}{\begin{equation}}
\newcommand{\ee}{\end{equation}}
\newcommand{\bsube}{\begin{subequations}}
\newcommand{\esube}{\end{subequations}}
\newcommand{\mean}[1]{\langle#1\rangle}

\def\i{{\rm i}}
\def\l{\ell}

\def\p{\partial}

\newif\ifusesec

\usesecfalse

\newif\ifuseappendix

\useappendixfalse

\usepackage{color}

\begin{document}

\title{Modeling the complete gravitational wave spectrum 
  of neutron star mergers}

\author{Sebastiano \surname{Bernuzzi}$^{1,2}$}
\author{Tim \surname{Dietrich}$^3$}
\author{Alessandro \surname{Nagar}$^4$}
\affiliation{$^1$TAPIR, California Institute of
  Technology, 1200 E California Blvd,Pasadena, California 91125, USA} 
\affiliation{$^2$DiFeST, University of Parma, and INFN Parma, I-43124
  Parma, Italy} 
\affiliation{$^3$Theoretical Physics Institute, University of Jena,
  07743 Jena, Germany} 
\affiliation{$^4$Institut des Hautes Etudes Scientifiques, 91440
  Bures-sur-Yvette, France}  
 
\date{\today}

\begin{abstract} 
In the context of neutron star mergers, we study the
gravitational wave spectrum of the merger remnant using numerical
relativity simulations. 
Postmerger spectra are characterized by a main peak frequency $f_2$
related to the particular structure and dynamics of the remnant hot
hypermassive neutron star. 
We show that $f_2$ is correlated with the tidal coupling
constant $\kappa^T_2$ that characterizes the binary tidal
interactions during the late-inspiral--merger. 
The relation $f_2(\kappa^T_2)$ depends very weakly on the  
binary total mass, mass-ratio, equation of state, and thermal
effects.  
This observation opens up the possibility of developing a model 
of the gravitational spectrum of \textit{every} merger 
unifying the late-inspiral and postmerger descriptions. 
\end{abstract}

\maketitle

\paragraph*{Introduction.--}

Direct gravitational wave (GW) observations of binary neutron stars 
(BNS) late-inspiral, merger and postmerger by ground-based GW
inteferometric experiments can lead to the strongest
constraints on the equation of state (EOS) of matter at supranuclear
densities~\cite{Damour:2012yf,Read:2013zra,DelPozzo:2013ala,Lackey:2014fwa,Agathos:2015uaa,Bauswein:2011tp,Clark:2014wua}. There 
are two ways to set such constraints\footnote{
  GW observations of BNS mergers can also constrain the source
  redshift~\cite{Messenger:2011gi,Messenger:2013fya}.}:   
(I) measure the binary phase during
the last minutes of coalescence using matched filtered searches~\cite{Damour:2012yf,DelPozzo:2013ala,Lackey:2014fwa,Agathos:2015uaa};
(II) measure the postmerger GW spectrum frequencies
using burst searches~\cite{Bauswein:2011tp,Clark:2014wua}.

Method (I) relies on the availability of waveform models that include 
tidal effects and are accurate up to
merger~\cite{Favata:2013rwa,Lackey:2014fwa,Agathos:2015uaa}.  
Here, ``up to merger'' indicates the end of chirping signal in
a precise sense that will be described below.  
Tidal interactions are significant 
during the late stages of coalescence at GW frequencies
$f_\text{GW}\gtrsim400$~Hz (for typical binary masses), and affect the phase
evolution of the binary. The zero-temperature EOS is constrained by
the measure of the quadrupolar tidal coupling constant $\kappa^T_2$ (or
equivalent/correlated parameters, e.g.~\cite{Lackey:2014fwa}) 
that accounts for the magnitude of the tidal 
interactions~\cite{Damour:2009wj,Damour:2012yf}. 

Combining results from numerical relativity and the effective-one-body (EOB)
approach to the general relativistic two-body
problem~\cite{Buonanno:1998gg,Buonanno:2000ef,Damour:2000we,Damour:2001tu}, 
one can show that the merger dynamics of \textit{every} 
irrotational binary is characterized by the value of
$\kappa^T_2$~\cite{Bernuzzi:2014kca}.
At sufficiently small separations, the relevant dependency 
of the dimensionless GW frequency on the EOS, binary mass, 
and mass-ratio is completely encoded in the tidal coupling
constant\footnote{%
  The spin dependence is 
  approximately linear for small spins aligned with the orbital
  angular momentum.}.
A tidal effective-one-body model compatible with numerical
relativity data up to merger was introduced in~\cite{Bernuzzi:2014owa}, 
but no prescription is available to extend the model to the postmerger.  

Method (II) relies on the high-frequency GW spectrum, and can, in
principle, deliver a measure independent on (I)~\cite{Clark:2014wua}.  
Binary configurations with total mass 
$M\leq M_\text{thr}\sim2.9M_{\odot}$ 
are expected to produce a merger remnant
composed of a hot massive/hypermassive neutron star. 
The merger remnant has a characteristic GW spectrum 
composed of a few broad peaks around $f_\text{GW}\sim1.8-4$~kHz. 
The key observation here is that the main peak frequencies
of the postmerger spectrum strongly correlate with properties
(radius at a fiducial mass, compactness, etc.) of a
zero-temperature spherical equilibrium star in an EOS-independent
way~\cite{Bauswein:2011tp,Bauswein:2012ya}. Thus, a measure of the
peak frequency 
constraints the correlated star parameter. Recently, there has been
intense research on this topic, and various EOS-independent relations were 
proposed~\cite{Bauswein:2011tp,Bauswein:2012ya,Hotokezaka:2013iia,Bauswein:2013jpa,Bauswein:2014qla,Takami:2014zpa,Bauswein:2015yca,Takami:2014tva}.  
Most of the relations are constructed for equal-mass
configurations and do not describe generic configurations for different 
total masses and mass-ratios, e.g.~\cite{Hotokezaka:2013iia,Bauswein:2015arXiv150308769B}. 
Additionally, the postmerger GW spectrum might be influenced 
in a complicated way by thermal effects, magnetohydrodynamical
instabilities and dissipative processes. 

In this paper we observe that the coupling constant $\kappa^T_2$ can
also be used to determine the main features of the postmerger GW
spectrum in an EOS-independent way and for generic binary configurations, 
notably also in the unequal-mass case.
The observation opens up the possibility of modeling 
the complete GW spectrum of neutron star mergers  
unifying the late-inspiral and postmerger descriptions.
Geometrical units $c=G=1$ 
are employed throughout this article, unless otherwise stated. 
We use $f$ for the spectrum frequencies and $\omega$ for the
instantaneous, time-dependent frequency.

\paragraph*{Numerical Relativity GW Spectra.---}
\label{sec:sim}

\begin{figure*}[t]
  \includegraphics[width=\textwidth]{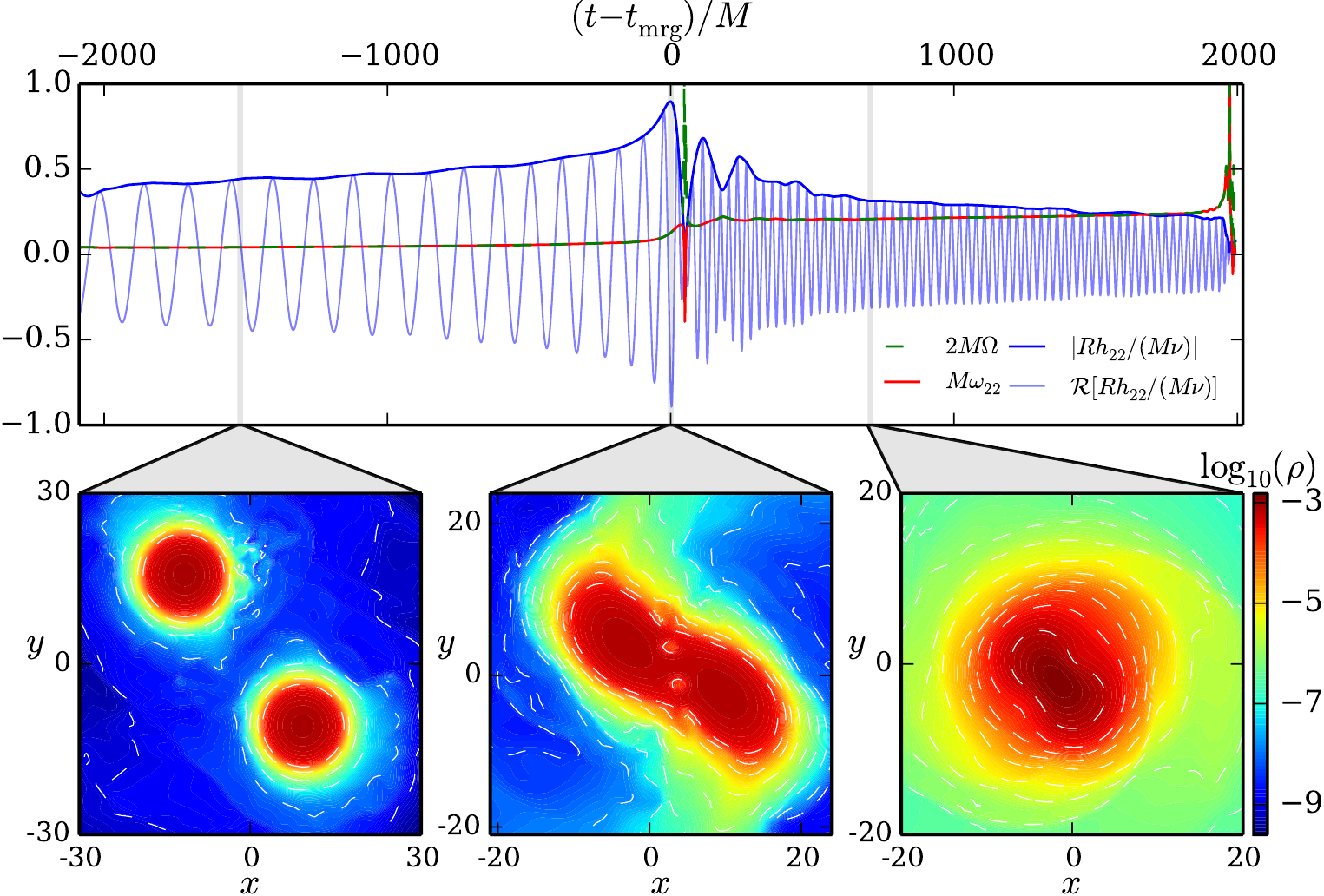} 
  \caption{Simulations of BNS and GWs.
    Top: real part and amplitude of the GW mode $R h_{22}/(\nu M)$
    and the associated dimensionless frequency $M\omega_{22}$ 
    versus the mass-normalized retarded time $t/M$ for a 
    fiducial configuration, H4-135135. 
    The signal is shifted to the moment of merger, $t_\text{mrg}$, 
    defined by the amplitude's peak (end of chirping).
    Also shown is (twice) the dynamical frequency $M\Omega=\p E_b/\p
    j\sim M\omega_{22}/2$.
    Bottom: Snapshots of $\log_{10}\rho$ on the orbital plane,   
    during the late inspiral (left), 
    at simulation time corresponding to $t_\text{mrg}$ (middle), 
    during the postmerger (right).}  
  \label{fig:hwaves}
\end{figure*} 

\begin{figure}[t]
   \includegraphics[width=0.5\textwidth]{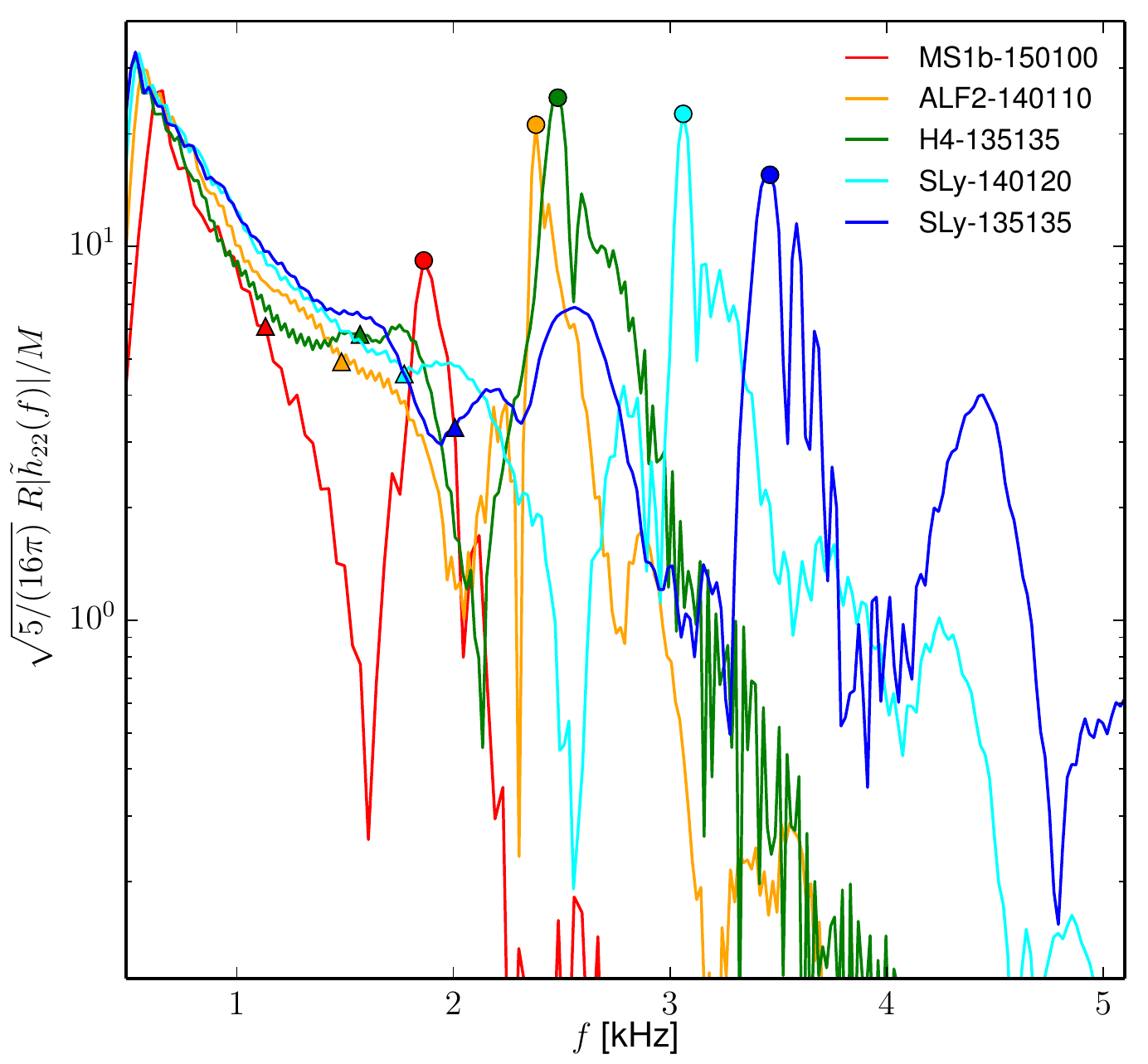}
   \caption{GWs spectra from BNS.
     The plot shows only a representative subset of the
     configurations of Table~\ref{tab:conf}. Triangles 
     mark frequencies $f_\text{mrg}$ corresponding to $t_\text{mrg}$, 
     circles mark $f_2$ frequencies.}
   \label{fig:hwaves_psd}
\end{figure}

\begin{table}[t]
  \centering    
  \caption{BNS configurations and data.
    Columns: name, EOS, binary total 
    mass $M$, mass ratio $q$, $f_2$ frequency in kHz,
    dimensionless $Mf_2$ frequency, 
    tidal coupling constant $\kappa_2^T$. 
    Configurations marked with~$^*$ are stable MNS.} 
 \begin{ruledtabular}  
  \begin{tabular}{ccccccc}        
    Name & EOS  & $M$ [$M_\odot$] & $q$ &$f_2$ [kHz]& $Mf_2$
    [$\times10^2$] & $\kappa_2^T$\\ 
    \hline
    SLy-135135    & SLy & 2.70  & 1.00  & 3.48  & 4.628  & 74  \\ 
    SLy-145125    & SLy & 2.70  & 1.16  & 3.42  & 4.548 & 75 \\ 
    ENG-135135    & ENG & 2.70  & 1.00  & 2.86  & 3.803 & 91  \\ 
    SLy-140120    & SLy & 2.60  & 1.17  & 3.05  & 3.906 & 96  \\ 
    MPA1-135135   & MPA1 & 2.70  & 1.00  & 2.57  & 3.418 & 115 \\ 
    SLy-140110    & SLy & 2.50  & 1.27  & 2.79  & 3.426 & 126 \\ 
    ALF2-135135   & ALF2 & 2.70  & 1.00  & 2.73  & 3.630 & 138 \\ 
    ALF2-145125   & ALF2 & 2.70  & 1.16  & 2.66  & 3.537 & 140 \\ 
    H4-135135     & H4 & 2.70  & 1.00  & 2.50  & 3.325 & 211 \\ 
    H4-145125     & H4 & 2.70  & 1.16  & 2.36  & 3.138 & 212 \\ 
    ALF2-140110   & ALF2 & 2.50  & 1.27  & 2.38  & 2.931 & 216 \\ 
    MS1b-135135$^*$   & MS1b & 2.70  & 1.00  & 2.00  & 2.660 & 290 \\ 
    MS1-135135$^*$    & MS1 & 2.70  & 1.00  & 1.95  & 2.593 & 327 \\ 
    MS1-145125$^*$    & MS1 & 2.70  & 1.16  & 2.06  & 2.740 & 331  \\ 
    2H-135135$^*$     & 2H & 2.70  & 1.00  & 1.87  & 2.561 & 439 \\ 
    MS1b-140110$^*$   & MS1b & 2.50  & 1.27  & 2.08  & 2.487 & 441 \\ 
    MS1b-150100$^*$   & MS1b & 2.50  & 1.50  & 1.87  & 2.303 & 461 \\ 
  \end{tabular}
\end{ruledtabular}  
 \label{tab:conf}
\end{table}

The numerical relativity data used in this work were previously
computed in~\cite{Bernuzzi:2014kca,Dietrich:2015iva}. 
In our simulations we solve Einstein equations
using the Z4c formulation~\cite{Bernuzzi:2009ex} and general
relativistic hydrodynamics~\cite{Font:2007zz}. Our numerical methods
are detailed
in~\cite{Dietrich:2015iva,Hilditch:2012fp,Bernuzzi:2012ci,Bernuzzi:2011aq,Thierfelder:2011yi,Brugmann:2008zz}. 
The binary configurations considered here are listed in
Table~\ref{tab:conf}. In the following we summarize the main features
of the GW radiation obtained by BNS simulations.    

We consider equal and unequal masses configurations, different total
masses, and a large variation of zero-temperature
EOSs parametrized by piecewise polytropic fits~\cite{Read:2008iy}. Thermal
effects are simulated with an additive thermal contribution in the 
pressure in a $\Gamma$-law form, $P_\text{th}= (\Gamma_\text{th}-1) \rho
\epsilon$, where $\Gamma_\text{th}=1.75$, $\rho$ is the rest-mass density
and $\epsilon$ the specific internal energy of the fluid,
see~\cite{Shibata:2005ss,Thierfelder:2011yi,Bauswein:2010dn}.   
The initial configurations are prepared in
quasicircular orbits assuming the fluid is irrotational. 

Initial data are evolved for several orbits, during 
merger and in the postmerger phase for $\gtrsim30$~milliseconds. 
A detailed discussion of the merger properties determined by different
EOSs, mass, and mass-ratio is presented
in~\cite{Bernuzzi:2014kca,Dietrich:2015iva}.  
The binary configurations in our sample do not promptly collapse to a 
black hole after merger, but form either a stable \textit{massive neutron
  star} (MNS) or an unstable \textit{hypermassive neutron star}
(HMNS), 
which collapses on a dynamical timescale
$\tau_\text{GW}\lesssim\mean{R}^4/\mean{M}^3\approx200$~ms~\cite{Baumgarte:1999cq}. 
Both HMNS and MNS remnants at formation are hot, differentially
rotating, nonaxisymmetric, highly dynamical two-cores structures,
e.g.~\cite{Shibata:2005ss,Stergioulas:2011gd}. 
   
The typical GW signal computed in our simulations is shown in
Fig.~\ref{fig:hwaves} for a fiducial configuration.
We plot the real part and amplitude of the  
dominant $\ell=m=2$ multipole of the $s=-2$ spin-weighted spherical
harmonics decomposition of the GW, $R(h_+-\i h_\times)=\sum_{\ell m}R h_{\ell
  m}\, {}_{-2}Y_{\ell m}(\theta,\phi)$, versus the retarded time, $t$. 
The figure's main panel also shows the
$\ell=m=2$ instantaneous and dimensionless GW frequency
$M\omega_{22}=M d\phi/dt$ where $\phi=-{\rm arg}(R h_{22})$. 
The bottom panels show snapshots of $\log_{10}\rho$
on the orbital plane, corresponding to three representative
simulations times. 

The waveform at early times is characterized by the well-known
chirping signal; frequency and amplitude monotonically increase in time. 
The GW frequency reaches typical values $\omega_\text{GW}=2\pi
f_\text{GW}\approx \omega_{22}\lesssim0.1-0.2/M$,
i.e.~$f_\text{GW}\sim 0.8-1.6$~kHz for  
a $M=2.7M_\odot$ binary.
The chirping signal ends at the amplitude peak, $\max|Rh_{22}|$, which
is marked in the figure by the middle vertical line. We formally define this
time as the \textit{moment of merger}, $t_\text{mrg}$, and
refer to the signal at $t>t_\text{mrg}$ as the postmerger signal.
The GW postmerger signal is essentially generated by the $m=2$ 
structure of the remnant, see bottom right panel of 
Fig.~\ref{fig:hwaves}. The frequency increases monotonically to 
$M\omega_{22}\sim0.2-0.5$ as the
HMNS becomes more compact and eventually approaches the collapse.
Assuming the remnant can be instantaneously approximated 
by a perturbed differentially rotating star~\cite{Stergioulas:2011gd}, 
the $f$-mode of pulsation is strongly excited at formation 
and it is the most efficient emission channel for GWs.   

The GW spectra are shown in Fig.~\ref{fig:hwaves_psd} for a
representative subset of configurations. Triangles mark 
frequencies $f_\text{mrg}$ corresponding to $t_\text{mrg}$. 
Circles mark the main postmerger peak frequencies $f_2\sim1.8-4$~kHz.
The small frequency cut-off is artificial and related to the small 
binary separation of the initial data; physical spectra monotonically 
extend to lower frequencies. 
From the figure one also observes that: 
(i)~there exists other peaks, expected by
nonlinear mode coupling or other hydrodynamical
interactions~\cite{Shibata:2005ss,Stergioulas:2011gd,Bauswein:2015yca}; 
(ii)~peaks are broad, reflecting the nontrivial time-evolution 
of the frequencies (see Fig.~\ref{fig:hwaves} and also the 
spectrogram in~\cite{Bernuzzi:2013rza});
(iii)~secondary peaks are present in
most of the configurations, their physical interpretation 
has been discussed
in~\cite{Stergioulas:2011gd,Takami:2014zpa,Bauswein:2015yca,Dietrich:2015iva}. 
We postpone the analysis of these features to future work.
In the following we focus only on the $f_2$ peak, which is the most
robust and understood feature of the GW postmerger spectrum.

\begin{figure*}[t]
  \includegraphics[width=\textwidth]{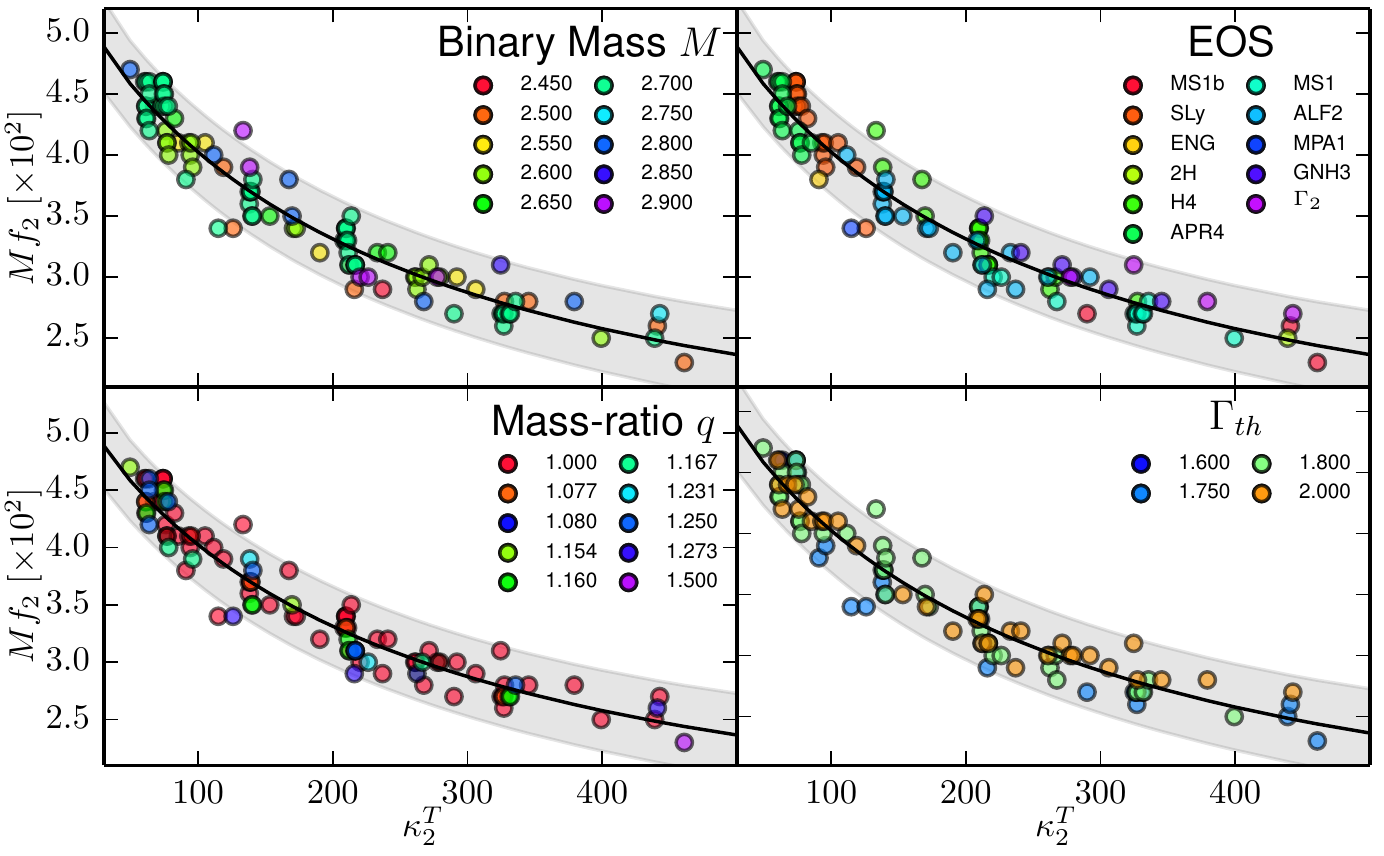}
  \caption{$Mf_2$ dimensionless frequency as a function of the tidal coupling
    constant $\kappa_2^T$. 
    Each panel shows the same dataset; the color code in
    each panel indicates the different values of binary mass (top left), EOS
    (top right), mass-ratio (bottom left), and $\Gamma_\text{th}$
    (bottom right).
    The black solid line is our fit (see Eq.~\eqref{fitmodel} and
    Table~\ref{tab:fit_coefs});  
    the grey area marks the 95\% confidence interval.} 
  \label{fig:f2_kappa}
\end{figure*}

\begin{table}[t]
  \centering    
  \caption{Fit coefficients of different quantities at $t_\text{mrg}$
    and of $Mf_2$ with the template in~\eqref{fitmodel}.}
  \begin{ruledtabular}  
  \begin{tabular}{lcccccc}          
    $Q(\kappa^T_2)$ & $Q_0$ & $n_1$ $[\times 10^2]$ & $n_2$ $[\times 10^5]$ & $d_1$ $[\times 10^2]$\\ 
    \hline
    $E_b^\text{mrg}$         & $-0.1201$     & +2.9905 & $-1.3665$  & $+6.7484$ \\ 
    $j^\text{mrg}$           &  $+2.8077$    & +4.0302 & +0.7538  & +3.1956 \\ 
    $M\omega_{22}^\text{mrg}$ & $+0.3596$     & +2.4384  & $-1.7167$ & $+6.8865$ \\ 
    $Mf_{2}$                & $+0.053850$   & $+0.087434$ & 0      & +0.45500 \\    
  \end{tabular}
\end{ruledtabular}
 \label{tab:fit_coefs}
\end{table}

\paragraph*{Characterization of the postmerger GW spectra.---}
Here, we show that $f_2$ correlates with the tidal coupling
constant $\kappa^T_2$ that 
parametrizes the binary tidal interactions and waveforms during 
the late-inspiral--merger. The relation $f_2(\kappa^T_2)$ depends very weakly on
the binary total mass, mass-ratio, and EOS. We use a large data sample
of 99 points including the data of~\cite{Hotokezaka:2013iia,Takami:2014tva}.

Let us first briefly summarize the definition of $\kappa^T_2$ and its
role in the \textit{merger} dynamics.

Within the EOB framework, tidal interactions are 
described by an additive correction $A^T(r)$ to the
radial, Schwarzschild-like metric 
potential $A(r)$ of the EOB Hamiltonian~\cite{Damour:2009wj}.
The potential $A(r)$ represents the binary interaction
energy. In order to understand its physical meaning, it is sufficient to
consider the Newtonian limit of the EOB Hamiltonian, 
$H_\text{EOB}\approx M c^2 + \frac{\mu}{2}\mathbf{p}^2 +
\frac{\mu}{2}(A(r)-1)
+ \mathcal{O}(c^{-2})$, 
where $\mu= M_A M_B/(M_A + M_B)$ is the binary reduced mass,
$\mathbf{p}$ the momenta, and 
$A(r) = 1 - \frac{2}{r} = 1 - 2 \frac{GM}{c^2 r_{AB}}$,  
with $r_{AB}$ the relative distance between the stars (constants $c$
and $G$ are re-introduced for clarity).
The tidal correction $A^T(r)$ is parametrized by a multipolar set of
relativistic 
tidal coupling constants $\kappa^{A,B}_{(\ell)}$, where $A,B$
label the stars in the binary~\cite{Damour:2009wj,Damour:2012yf}. 
The leading-order contribution to $A^T(r)$ is proportional to the quadrupolar
($\l=2$) coupling constants, 
$\kappa^A_2 = 2 k^A_2 \left(X_A/C_A\right)^5 M_B/M_A$ 
where $M_A$ is the mass of star $A$, $C_A$ the compactness, $X_{A}=M_{A}/M$,
and $k_{2}^{A}$ the $\ell=2$ dimensionless Love
number~\cite{Hinderer:2007mb,Damour:2009vw,Binnington:2009bb,Hinderer:2009ca}.
The total $\l=2$ coupling constant is defined as
$\kappa^T_2=\kappa^A_2+\kappa^B_2$, and can be written as
\be
\label{kappaT}
\kappa^T_2 = 2 \left( \frac{q^4}{(1+q)^5} \frac{k_2^A}{C_A^5}    +
\frac{q}{(1+q)^5} \frac{k_2^B}{C_B^5} 
\right)  \ ,
\ee
assuming $q=M_A/M_B\geq1$. The leading-order term of the tidal
potential is simply $A^T(r)=-\kappa^T_2 r^{-6}$. 

A consequence of the latter expression for $A^T(r)$ is that 
the merger dynamics is essentially determined by the value 
of $\kappa^T_2$~\cite{Bernuzzi:2014kca}.
All the dynamical quantities develop a nontrivial dependence on
$\kappa^T_2$ as the binary interaction becomes tidally dominated.
The characterization of the merger dynamics via $\kappa^T_2$ is
``universal'' in the sense that it does not require any other
parameter such as EOS, $M$, and $q$. (There is, however, a
dependency on the stars spins.) 
For example, at the reference point $t_\text{mrg}$, the corresponding binary reduced
binding energy $E_b^\text{mrg}$, the reduced angular momentum
$j^\text{mrg}$, and the GW frequency $M\omega_{22}^\text{mrg}$  can 
be fitted to simple rational polynomials~\cite{Bernuzzi:2014kca}
\be
\label{fitmodel}
Q(\kappa^T_2)=Q_0\frac{1+n_1 \kappa^T_2+n_2 (\kappa^T_2)^2}{1+d_1 \kappa^T_2}
\ ,
\ee
with fit coefficients $(n_i,d_i)$ given in Table~\ref{tab:fit_coefs}.

In view of these results, it appears natural to investigate the
depedency of the \textit{postmerger} spectrum on $\kappa^T_2$.

Our main result is summarized in Fig.~\ref{fig:f2_kappa}, which
shows the postmerger main peak dimensionless frequency $Mf_2$ 
as a function of $\kappa_2^T$ for a very large sample of binaries. 
Together with our data we include those tabulated
in~\cite{Hotokezaka:2013iia,Takami:2014tva}.  
The complete dataset spans the ranges $M\in[2.45 M_\odot,2.9 M_\odot]$, 
$q\in[1.0,1.5]$, and a large variation of EOSs.
The peak location is typically determined
within an accuracy of $\delta f\sim\pm0.2$~kHz, 
see also~\cite{Bauswein:2012ya}. 
Each of the four panels of Fig.~\ref{fig:f2_kappa} shows the same
data; the color code in 
each panel indicates different values of $M$ (top left), EOS
(top right), $q$ (bottom left), and $\Gamma_\text{th}$ (bottom right).
The data correlate rather well with $\kappa_2^T$. 
As indicated by the colors and different panels, the scattering of the
data does \emph{not} correlate with variations of $M$, EOS, $q$,
$\Gamma_{th}$. 
The black solid line is our best fit to Eq.~\eqref{fitmodel}, where we set
$n_2=0$ and fit also for $Q_0$, see Table~\ref{tab:fit_coefs}.
The fit 95\% confidence interval is shown
as a gray shaded area in Fig.~\ref{fig:f2_kappa}.  

We argue that the observed postmerger correlation with $\kappa^T_2$ is a 
direct consequence of the merger universality. Although an
analytical/approximate description of the postmerger dynamics is
not available, the gauge-invariant $E_b(j)$ curves contain, in 
analogy to the merger case, significant information about the system
dynamics~\cite{Bernuzzi:2014owa}. Specifically, we interpret $E_b(j)$
as being generated by some Hamiltonian flow that continuously connects
merger and postmerger.  
In terms of this Hamiltonian evolution, the values
$(E_b^\text{mrg}(\kappa^T_2),j^\text{mrg}(\kappa^T_2))$ provide initial 
conditions for the dynamics of the MNS/HMNS; it is  then plausible to
assume that the postmerger correlation follows from these initial
conditions by continuity.
In order to assess this conjecture, we define the frequency given 
by the equation $M\Omega=\p E_b/\p j$, notice that
$\Omega^\text{mrg}=\Omega^\text{mrg}(\kappa^T_2)$, and show that
$\Omega$ is the relevant dynamical frequency for both
inspiral-merger and postmerger.
Recalling that the standard quadrupole formula predicts that 
a generic source with $m=2$ geometry and rotating at frequency
$\Omega$ emits GWs at a frequency $2\Omega$, 
we plot the latter in Fig.~\ref{fig:hwaves} and indeed observe that it
corresponds to the main emission channel $\omega_{22}$ during the
whole evolution.
In practice, the gauge-invariant $\Omega$ can be interpreted as the 
orbital frequency during the inspiral, and the angular frequency of
the MNS/HMNS during postmerger.
Furthermore, since merger remnants from larger $\kappa^T_2$ binaries 
are less bound and have larger angular momentum support at formation, 
$\Omega(\kappa^T_2)$ (so $f_2$) 
must be a monotonically decreasing function of $\kappa^T_2$, 
which is what one can observe in Fig.~\ref{fig:f2_kappa}. 

The frequency evolution is also expected to depend on angular momentum
dissipation due to magnetic fields instabilities,
e.g.~\cite{Ciolfi:2011xa,Zink:2011kq,Kiuchi:2014hja},
cooling and shear viscosity~\cite{Baumgarte:1999cq}. However, the available
literature indicates these physical effects are negligible in first
approximation, and we argue that they might result in frequency shifts
$\Delta f_2\lesssim\delta f_2$. 
The stars rotation can instead play a relevant role 
via spin-orbit coupling effects: stars with dimensionless spin
parameters $\gtrsim0.05-0.1$ can give  
frequency shifts $\gtrsim\delta f_2$~\cite{Bernuzzi:2013rza}.

\paragraph*{Outlook.---}
\label{sec:conc}

The result of this work, coupled with the modeling of the merger process given
in~\cite{Bernuzzi:2014kca,Bernuzzi:2014owa}, indicates the possibility
to model the  
late-inspiral-merger-postmerger GW spectrum in a consistent way 
using $\kappa^T_2$ as main parameter. In particular, an accurate
late-inspiral-merger GW 
spectrum is given by a suitable frequency-domain representation,
$\tilde{h}(f)=A(f)\exp{[-\i\Psi(f)]}$, of the waveform
of~\cite{Bernuzzi:2014owa}.
The leading-order tidal contribution of such a spectrum reads
$\Psi^T(f)=-39/4\, \kappa^T_2 \, x^{5/2}$ with  
$x(f)\propto f^{2/3}$; see~\cite{Damour:2012yf} for $\Psi^T(f)$ at 2.5
post-Newtonian order.  
A simple template for the postmerger spectrum for binaries with 
$M\leq M_\text{thr}$ is then given by a single-peak-model
and our fit for $f_2$. The precise construction of such complete spectrum
will be subject of future work. As mentioned in our discussion,
it will be particular important to  
include spin effects, e.g.~\cite{Bernuzzi:2013rza,Agathos:2015uaa}.

The performance of the proposed model in a GW data-analysis
context will be carefully evaluated in a separated study. In this
respect, we suggest that an optimal strategy to constrain the EOS
could be combining the late-inspiral measurement of type (I) with
measurement of type (II). The inclusion of the postmerger model might
lead to an improved estimate of $\kappa^T_2$, for the same number of observed
events~\cite{DelPozzo:2013ala,Lackey:2014fwa,Agathos:2015uaa}.

\paragraph*{Acknowledgments.---}
We thank Andreas Bauswein, Bernd Br\"ugmann, Tibault Damour, Sarah
Gossan, Tjonnie Li, David Radice, Maximiliano Ujevic, Loic Villain for
comments and 
discussions.      
This work was supported
in part by DFG grant SFB/Transregio~7 ``Gravitational Wave Astronomy''  
and the Graduierten-Akademie Jena. 
S.B. acknowledges partial support from the National
Science Foundation under grant numbers NSF AST-1333520, PHY-1404569, and
AST-1205732.  T.~D. thanks IHES for hospitality during the development of
part of this work.
Simulations where performed at LRZ (M\"unich) and at JSC (J\"ulich).

\bibliography{../../../Refs/sb_refs.bib}

\end{document}